\documentstyle[prb,aps,epsf,multicol]{revtex}
\begin{document}
\draft 
\title
{Atomistic modelling of large-scale metal film growth fronts}
\author{U. Hansen,$^{1}$ P. Vogl,$^{1}$ and Vincenzo Fiorentini$^{1,2}$}
\address{{\it 1)\,} Physik-Department and Walter Schottky Institut,
Technische  Universit\"{a}t  M\"{u}nchen, D-85748 Garching, Germany\\
{\it 2)\,} Istituto Nazionale per la Fisica della Materia 
and Dipartimento di Fisica, Universit\`a di Cagliari, Italy\\
}
\date{\today}
\maketitle
\begin{abstract}
We present simulations of metallization morphologies  under
ionized sputter deposition conditions, obtained by
a new theoretical approach. By means of
molecular dynamics simulations using a
 carefully designed  interaction potential, we
 analyze the surface  adsorption, 
reflection, and etching reactions 
taking place during Al physical vapor deposition,
and calculate their relative probability. These probabilities are
then employed  in a feature-scale cellular-automaton simulator, which
produces calculated film morphologies in excellent agreement with
scanning-electron-microscopy data on ionized sputter deposition.   
\end{abstract}

\vspace{-0.3cm}
\begin{multicols}{2}
In this Letter we present results of a hierachy of theoretical models
developed to describe the growth of metal thin films. Atomic-scale
molecular dynamics  (MD) and a  feature-scale
cellular-automaton simulator are combined to yield
  realistic simulations of film growth during physical vapor
 metallization of contact vias typical of semiconductor device
technology. The MD simulations, accounting in full for the
microscopic many-atom dynamics, are used to predict the
reaction rates of the processes relevant in physical
vapor deposition;  the cellular automaton 
simulator incorporates the reaction rates thus obtained, enabling us
to predict and understand the topography of $\mu$m-scale film fronts,
 and their relationship
to substrate geometry, incident beam energy, and angular beam 
distribution. Our approach yields a consistent and computationally
efficient scheme to predict the topography of metal films on
arbitrarily-shaped substrates. A comparison
with scanning-electron-microscopy data on sputter-deposited Al-covered
trenches demonstrates an  excellent level of agreement between  
theory and experiment. 

Our theoretical approach proceeds in three steps, namely
{\it (a)} a classical interatomic interaction potential for Al is
developed,
({\it b}) reaction rates for Al atoms incident  on Al surfaces are
 calculated therewith in a MD simulation, and
({\it c}) a cellular automaton is developed and
employed to simulate   $\mu$m-scale film fronts,   using the reaction
rates extracted from MD. In the following, we analyze this procedure
and present the relevant results for each step.	

\paragraph*{ Interatomic potential -- }
Previously developed classical interatomic
potentials \cite{daw,cai,hoagland,baskes,mei,ercolessi} for Al-Al
interactions have mostly focused on bulk and molecular properties. The
 Ercolessi-Adams potential \cite{ercolessi} is a partial  exception 
since,
besides reproducing quantitatively the bulk and elastic properties of
Al, it also yields interlayer relaxations \cite{ercolessi} in
good agreement with experiment.\cite{kamm}  Using this potential, we
obtain surface energies in excellent agreement with {\em ab initio}
calculations.\cite{deyirmenjian} We  thus chose this model as a
starting point to develop a new classical many body potential for
Al. Our own potential is carefully designed to reproduce the
properties of Al aggregates  in a wide range of bonding configurations
(from bulk  Al to Al surfaces and Al molecules), with special regard
to surface properties. Its functionalities  significantly extend those
of previous embedded atom models,
 while maintaining a high level of accuracy in  reproducing Al bulk
properties   and surface formation
energies.\cite{ercolessi} {\it First}, we introduced   an exponential
repulsive pair potential \cite{abrahamson,nota}  to account
for the short-range interaction of  Al  atoms with kinetic energies
exceeding 10 eV; this is
a key requirement, as the kinetic energies of deposited atoms 
reach over 150 eV during ionized physical vapor deposition.
{\it  Second},  the embedding function has been  readjusted  to
reproduce observed properties of low-density Al structures; after
these changes, we obtained a significantly improved agreement of
several reference quantities \cite{compar} in comparison to {\it ab inito}
results\cite{stumpf,feibelman}
 and/or experiment.\cite{djugan}  {\it Third}, we
introduced a 5th-order polynomial cutoff function, smoothly cutting
off the interaction range of our potential 
 at an interatomic distance of $5.56$ \AA, slightly larger than
third-nearest neighbor distance in bulk Al.

As a further test of our potential, we calculated the
 barriers for homodiffusion on the low-index faces of fcc Al.
Specifically,  we considered hopping on Al (110)
along and orthogonal to the [1$\overline{\rm 1}$0]-oriented rows,
 hopping on Al (100), and concerted exchange on Al (100).
 The general level of agreement with {\it ab initio}
density-functional calculations  \cite{stumpf,feibelman} is very good
(as can be seen  in Table \ref{diffusion_barriers}), giving
us further confidence  in the accuracy of our potential.

\vspace{-0.1cm}
\narrowtext\begin{table}[th]
\caption{Comparison of selected hopping and exchange  diffusion 
 barriers on low-index Al
surfaces obtained with the present model and in {\it ab initio}
calculations. Al (111) is also included for completeness.}
\label{diffusion_barriers} 
\begin{tabular}{lcc}
               & This work & {\it Ab initio} \\ \hline
Al (111) hopping       & 0.04 & 0.04\tablenotemark[1] \\
Al (100) hopping      & 0.60 & 0.68\tablenotemark[1],0.65\tablenotemark[2]\\
Al (100) exchange    & 0.50&  0.35\tablenotemark[1] \\
Al (110) $\bot$ hopping   & 1.13 & 1.06\tablenotemark[1] \\
Al (110) $\|$ hopping & 0.30 & 0.60\tablenotemark[1] \\
\end{tabular}
\tablenotemark[1]{Reference \onlinecite{stumpf}};\,
\tablenotemark[2]{Reference \onlinecite{feibelman}} \end{table}

\paragraph*{Reaction rates from molecular dynamics -- }
In the second step of our approach,  reaction probabilities were
calculated in classical molecular dynamics simulations using our 
Al interaction potential. In particular we determined,
as  a function of the energy and off-normal angle of incident Al
atoms, the probability of three processes: adsorption, reflection, 
and etching (in the latter, the incoming atom's impact on the 
surface causes the kick-out of one or more substrate atoms).
 Supercells containing 1320 atoms arranged in 10 atomic layers 
are employed; cell dimensions are chosen so as to avoid
artifacts of the in-plane periodicity.
The starting configuration is chosen to be a (111) surface, 
the one Al surface with the lowest formation energy. All atomic
 coordinates are allowed to evolve  dynamically, except those of the
two bottom layers of the supercell. The surface temperature is set at
450 K (i.e about 1/2 of the melting temperature, and 20 \% larger than
the bulk Debye temperature).

We start our simulations with the incident Al atom placed  
outside the interaction range of the surface. Its initial  kinetic
energy is set in the range of 25 to 125 eV, and its starting
angle off the  surface normal in the range  $0^{\circ}$ to
$60^{\circ}$, which corresponds to typical ionized physical vapor
deposition conditions. The trajectories of the incident atom, and of 
any other atom which may be etched away from the surface upon impact,
are then monitored until either  a certain time span has elapsed, or
the outcoming atoms (in the case of reflection or etching) have
traveled  a distance of 10 \AA\, away from the surface. Analyzing 200
trajectories per incident energy and angle, we collected a
statistically significant sample of well-defined adsorption,
reflection, and etching  events. The relative probability  of the
corresponding process is calculated as the ratio of the number of
events of each kind to the total number. The typical  statistical
error in the reaction probabilities thus determined is below 5 \%.  

\begin{figure}[th]
\begin{center}
\epsfclipon
\epsfxsize=8.5cm
\narrowtext
\centerline{\epsffile{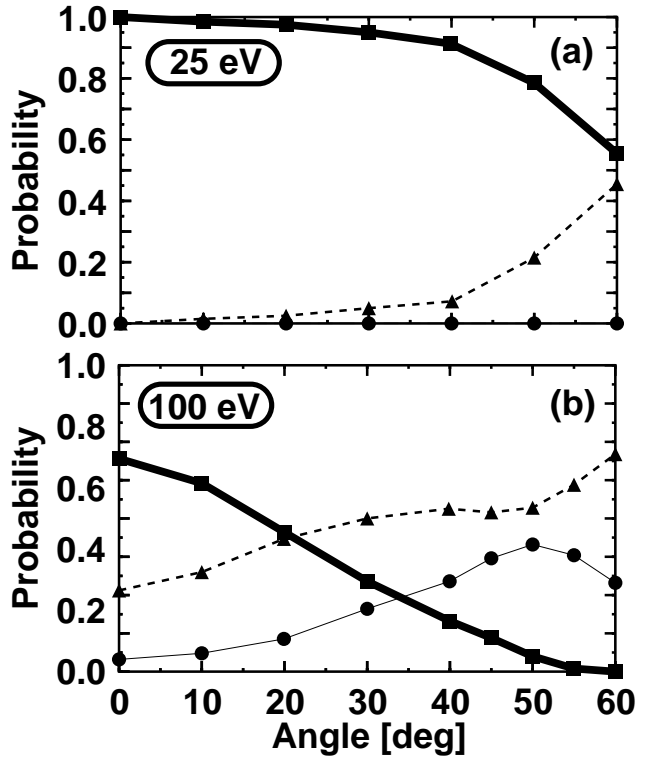}}
\hspace{0.1cm}
\caption{\label{figure1}
Reaction probabilities for Al atoms
impinging on Al(111)  with a kinetic energy of 25 eV
(panel (a)) and 100 eV (panel (b)) as a function of
the off-normal  angle. The processes considered are
adsorption (thick solid line, solid squares), 
reflection (dashed line, solid triangles), and etching (thin solid
line, solid circles).}
\end{center}
\vspace{-0.5cm}
\end{figure}

The behavior of reflection, adsorption, and etch rates as  a function
of off-normal angle are summarized in Fig. \ref{figure1} for two 
representative incident kinetic energies, namely  25 eV in panel (a),
and 100 eV in panel (b).  At low energies (panel (a)) the adsorption
probability (thick solid line, solid squares) decreases from near 
unity for angles below $20^{\circ}$ to approximately 1/2 for angles in
the range of $60^{\circ}$. This  decrease is compensated by a
corresponding increase in the reflection probability (dashed line,
solid triangles).  At  low  energy, etch processes  (thin solid
line, solid circles) are negligible at all angles.
Increasing  the incident kinetic energy to 100 eV, we find
considerable changes in the  relative reaction probabilities. Even at
small angles, the reflection probability is non-vanishing; the
adsorption probability is correspondingly reduced, and is now below
0.7 at all angles. More interestingly,  the etching probability is
always non-zero, it  reaches a  maximum of 0.4 at $50^{\circ}$, then
decreases as 
near-grazing angles are  approached. For small deviations from normal
incidence, the etch rate initially raises, since the
probability of a surface atom to gain momentum directed away from the
surface increases when the incoming atom arrives at an oblique angle
at the surface. At large angles of incidence the etch rate drops
 because of
the competing  specular reflection events.

By adjusting the bias voltage between sputter source and deposition
target  during ionized physical vapor deposition, both  the energy
regimes just discussed  are experimentally accessible. It is
expected that they lead to rather  different trench topographies, 
which we simulate with a cellular-automaton technique. 

\paragraph*{Feature-scale cellular-automaton simulator -- }
In the final step of our approach, we have developed a two-dimensional
cellular automaton to model the growing film front on a $\mu$m
scale. The automaton accounts for the effects of flux shadowing,
adsorption, reflection, etching, and surface diffusion. 
The simulated structure is represented in  cross section by a
two-dimensional grid. Each grid cell represents an Al atom,
and is assumed to have a physical length of 2.5 \AA\, (the
effective atomic diameter in Al bulk): thus,  for example,
a 1-$\mu$--wide structure will be described by 4000 grid cells 
across.

The atoms are serially and independently emitted from the sputter
source far above the surface according to a pre-determined angular and
energetic distribution, and move on a straight trajectory, determined
also by the applied source-target bias,  until they strike the growing
film front. Interactions in the gas phase are neglected in view of the
the low pressure (typically a few tenths of mTorr) and   the resulting
long mean free path typical of sputter deposition.  Spontaneous
desorption is also negligible in all the conditions considered. 

The impact angle and
energy of the atoms hitting the surface determine which process is
activated upon impact.\cite{angle-note}  This may be any one of the
three  (adsorption, reflection, etching) whose rates have been
previously calculated via MD simulations. If the atoms are
reflected, or an etch 
process takes place, the path of the corresponding atoms is
further traced until they hit the film surface for a second time. The
three basic processes can then  take place again, and so forth. Finally, 
the atoms get either adsorbed, or escape back into the gas phase.

Adsorption present some additional complications.
The local diffusion and accomodation mechanism upon adsorption is a key
ingredient in deposition models.  A reasonable assumption is
that incoming  atoms are accomodated at a (local)
minimum energy site within one diffusion length from the landing
site,  the diffusion length being determined by the surface
temperature and morphology, and  the deposition rate. In the cases of
interest here, the problem of determining an effective diffusion
length for the adsorbed atoms is quite formidable for several
reasons. On the one hand, one expects low effective diffusivity due
to the very high experimental  growth rate (not well controlled, but
in the order of 0.5 $\mu$m/minute, or roughly 40 ML/sec
\cite{hamaguchi,westwood}), and also because
 the growing surface rapidly looses its low-index
character becoming essentially disordered.  On the other hand,
collision energies are large, and (although  energy transfer
mechanisms  on rough surfaces are largely unknown) one may expect
multi atom events and transient mobility effects to increase the
effective  diffusivity. We aim at using parameters 
that  maximize the coordination of the new particle, with the
constraint that the 
impact-to-final site distance is minimized (so that the
 local film curvature is minimized\cite{simbad}).

In practice, we set the following criteria for ({\it a}) the maximum
diffusion distance, and  ({\it b}) the nature of a minimum-energy
site: ({\it a}) The
{\it  maximum} diffusion distance is taken to be $d_{\rm max}$ =
 $d_{\rm max}^{\rm thr}$ = 5 grid spacings = 12 \AA.
All sites within this distance from the landing site are analyzed;
the chosen final site is the one {\it closest} to the landing site,
among those with the highest local coordination (to be defined below,
 point ({\it b)}). Because of this, the effective  diffusion length 
is actually rather smaller than the maximum value  $d_{\rm max}$.
Our specific choice of $d_{\rm max}$ may be simply be regarded as
 a calibration factor designed to avoid  dendritic
structures or very flat surfaces which are not observed in the
experimental regime of interest here. However, we find
a rather sharp change in the   dependence of the surface roughness  on
$d_{\rm max}$,  from rather weak for  $d_{\rm max} > d_{\rm max}^{\rm
thr}$    
to strong for  $d_{\rm max} < d_{\rm max}^{\rm thr}$.\cite{asym-note}
 Thus our value $d_{\rm max}^{\rm thr}$ is
effectively a crossover threshold between the two regimes mentioned.
	
({\it b}) -- A minimum-energy site is defined in terms of its local
coordination\cite{methfessel}  as follows: for each candidate
site within the preset maximum distance  from the landing site (point
({\it a})) we calculate the number of atoms contained in a circle 
center at the candidate site and with a radius of 7 grid spacings. This
criterion amounts to selecting the site with maximum {\it average}
coordination; while similar to that of highest first neighbor  
coordination,  contrary to the latter  it avoids pathological choices
such as high coordination sites on highly ramified
structures.

\begin{figure}[ht]
\epsfclipon
\epsfxsize=8.5cm
\narrowtext
\centerline{\epsffile{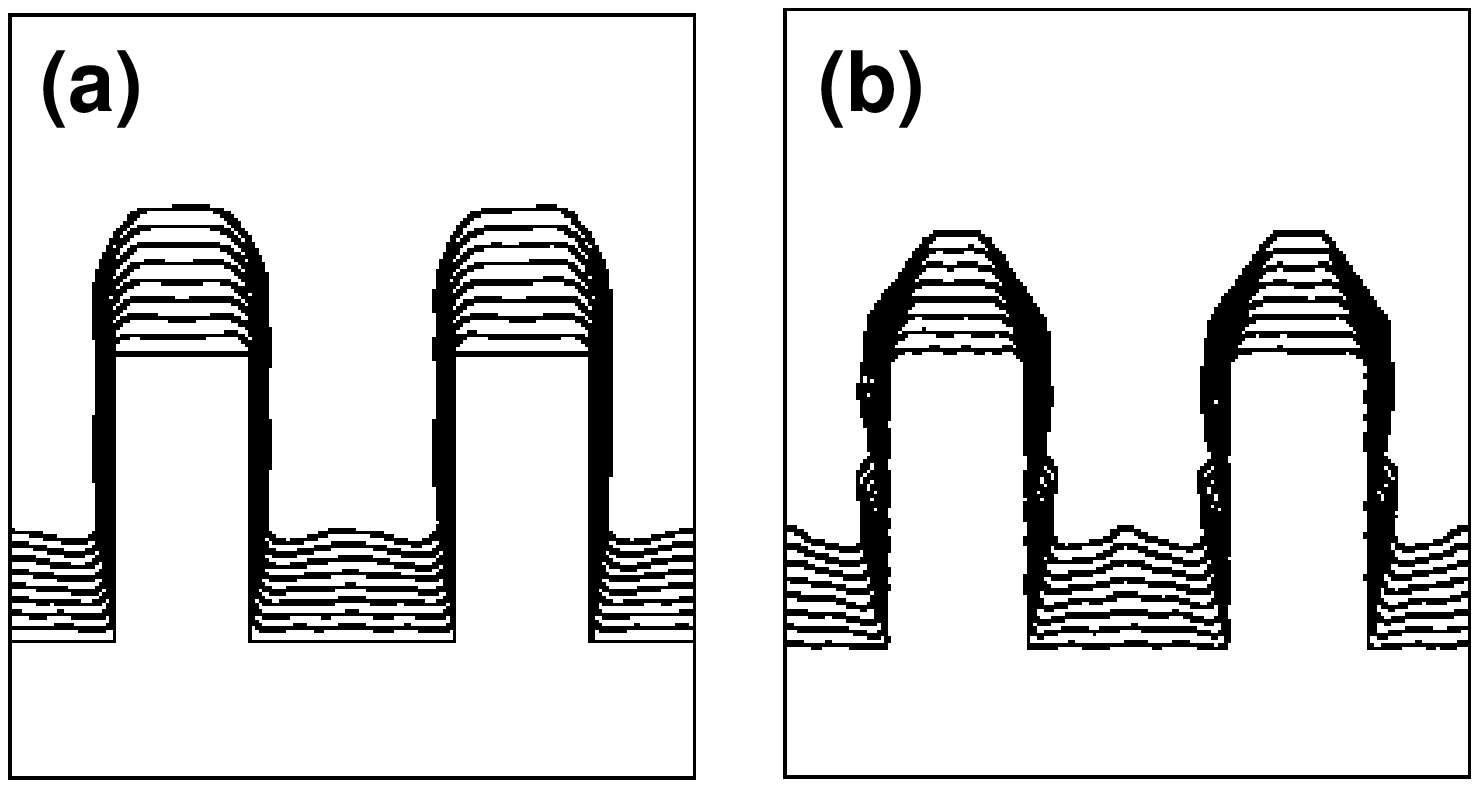}}
\epsfxsize=8.5cm
\narrowtext
\centerline{\epsffile{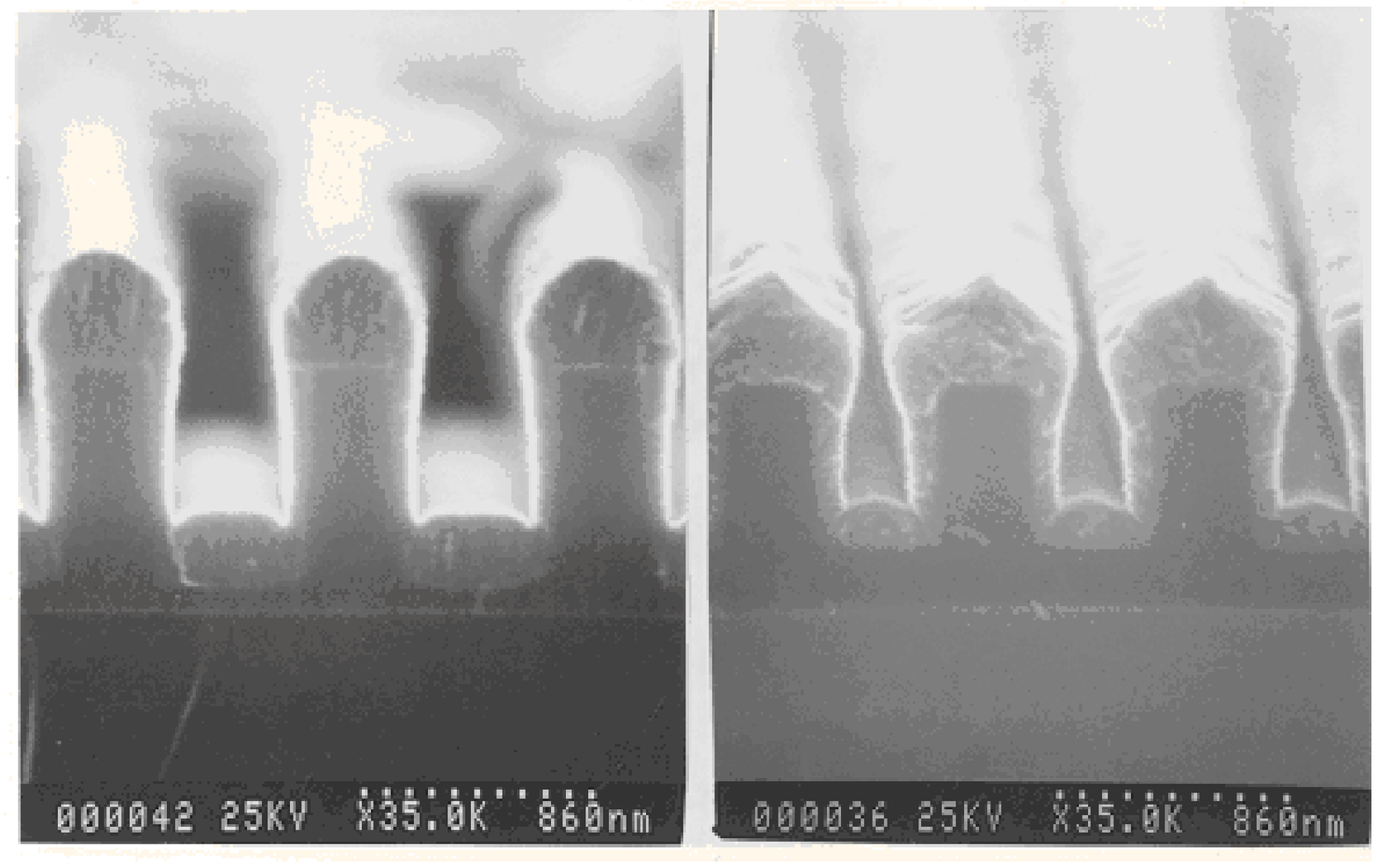}}
\hspace{0.1cm}
\caption{Film morphologies on trench structures 
predicted for different ionized magnetron
sputtering conditions compared to 
experiments.\protect\cite{hamaguchi}
In panel (a) we set
80\% ionization, and  10 V bias, in panel (b) 80\% ionization, and 80
V bias. Below, a SEM micrograph with  experimental results.}
\label{figure2}
\end{figure}

\paragraph*{Results of topography simulations -- }
Figure \ref{figure2} depicts trench topographies  predicted by our
model for different deposition conditions, compared to
scanning-electron-microscope pictures taken in similar 
conditions.\cite{hamaguchi} The structure size is 1.2 $\mu$m across.
In the cellular automaton simulation, the emitted atom energies 
are picked from  a Thompson distribution centered at 3 eV, as
suggested experimentally.\cite{westwood,dullini} The initial angles
of the non-ionized atoms are chosen from a collimated cosine
distribution with a maximum off-normal angle of $40^{\circ}$. To
mimic the experimental conditions, we assume that 80\% of the emitted
Al atoms get one-fold ionized; for these atoms, the trajectory and
impact energy  change according to the applied source--target bias. 

We now describe our results for different values of the
 the sputter source-to-target bias corresponding to the low and
high energy regimes identified in our MD calculations.
The results in panel (a) of
Fig. \ref{figure2} were obtained setting 
 the bias  to 10 eV. In agreement with experiment (also displayed in
Fig.2), we predict a 
 film growth front of rounded shape on top of the feature and, due
to geometric shadowing, a reduced  film thickness at the bottom of 
the trench. The pile-up at the center of  the trench  
is  not only of  geometric origin, but also partly due to 
reflections of atoms impinging on the trench sidewalls.
In panel  (b) of Fig. \ref{figure2} we report results obtained
with a bias of 80 eV. Our calculations
predict, in accord with experiment, a roof-like structure on top of
the feature. This structure is due to the preferential etching at
angles of $50^{\circ}$  (see Fig. \ref{figure1} (a)) 
which leads to a lower deposition rate on the roof-like structure.  
Simulations of structures scaled down in size by a similarity
 transformation,\cite{bs} (i.e. with the same geometry, relative sizes,
and  aspect ratios)
produced very  close results, suggesting that the
profiles obtained are largely self-similar at these length 
scales.\cite{bs}

In conclusion, we have demonstrated the viability of accurate
simulations of mesoscopic thin film morphology based on atomic-scale
simulations. We performed detailed theoretical calculations of 
the probabilities for surface reactions taking place during ionized
physical vapor deposition conditions, and combined these predictions
with $\mu$m-scale  cellular-automaton simulations. 
Our molecular dynamics calculations revealed strongly energy
dependent adsorption, reflection, and etch rates (the latter 
exhibiting  a distinct maximum for high incident kinetic energies
 at $\sim 50^{\circ}$). We were able to predict topographies of metal
films deposited on trench structures under different ionized physical
vapor deposition conditions in remarkable agreement with experiment.
Our results represent a major step ahead over earlier thin film growth
models  based on rate equations.\cite{friedrich}  which  did not
incorporate beam-energy--dependent surface reactions on an atomistic scale. 

We gratefully acknowledge financial support by Siemens AG. 
We thank Dr.\ A.\ Spitzer and 
Dr.\ A.\ Kersch for valuable guidance throughout the project,
and Dr. Paolo Ruggerone for a critical reading. V.F. was supported by
the Alexander von Humboldt-Stiftung during his stay at the Walter Schottky
Institut.

\narrowtext

\end{multicols}

\begin{thebibliography}{10}


\bibitem{daw}
 M. S. Daw and M. I. Baskes, Phys. Rev. Lett.
{\bf 50}, 1285 (1983);  Phys. Rev. B {\bf 29},
6443 (1984).

\bibitem{cai} J. Cai and Y. Y. Ye, Phys. Rev. B {\bf 54}, 8398 (1996).

\bibitem{hoagland} R. G. Hoagland, M. S. Daw, S. M. Foiles, and
M. I. Baskes, J. Mater. Res. {\bf 5}, 313 (1990). 

\bibitem{baskes} M. I. Baskes, Phys. Rev. B {\bf 46}, 2727 (1992).

\bibitem{mei} J. Mei and J. W. Davenport, Phys. Rev. B {\bf 46}, 21 (1992).

\bibitem{ercolessi} F. Ercolessi and J. B. Adams, Europhys. Lett. {\bf
26}, 583 (1994). 

\bibitem{kamm} G. N. Kamm and G. A. Alers, J. Appl. Phys. {\bf 35},
327 (1964).

\bibitem{deyirmenjian} V. B. Deyirmenjian, V. Heine, M. C. Payne,
V. Milman, R. M. Lynden-Bell, and M. W. Finnis, Phys. Rev. B {\bf 52},
15191 (1995).

\bibitem{abrahamson} A. A. Abrahamson, Phys. Rev. {\bf 178}, 178
(1969).

\bibitem{nota}
This potential is  V($r$) = $A\, \exp(-B r) - C$,
with $A=7255.44$ eV, $B=4.42085$ \AA$^{-1}$, and $C=1.04897$ eV.

\bibitem{compar}
The  binding energy and vibrational energy of
Al$_2$, the (111) homodiffusion barrier, 
the adsorption  energy per atom on Al (111),
and the binding energy of Al$_2$ on Al (111) with respect to isolated
adatoms are  --1.4 eV, 290 cm$^{-1}$, 0.04 eV,
2.42 eV, and 0.50 eV. They compare very well with
the reference values (all {\it ab initio} values 
from Ref. \onlinecite{stumpf} unless otherwise
specified)  --1.4 eV, 284 cm$^{-1}$ (experimental, Ref.
 \onlinecite{djugan}), 0.04 eV, 3.06 eV, and 0.58 eV.
The LDA adsorption energy  is larger than ours due to the known LDA 
overbinding, but the two values agree well when expressed in
percentage of the bulk binding  energy ($\sim$ 73 \%).


\bibitem{stumpf} R. Stumpf and M. Scheffler, Phys. Rev. Lett. {\bf
72}, 254 (1994);  Surf. Sci. {\bf 307}, 501 (1994);
Phys. Rev. B {\bf 53},4958 (1996). 


\bibitem{feibelman} P. J. Feibelman, Phys. Rev. Lett. {\bf 65},
729 (1990). 

\bibitem{djugan} M. F. Cai. T. P. Djugan, and V. E. Bondybey,
Chem. Rev. Lett. {\bf 155}, 430 (1989). 

\bibitem{angle-note}
The local normal at the impact site is defined as the 
direction connecting the impact point to the center of mass of the
atoms contained in a circle centered on the impact point and with
radius of 10 grid spacings.


\bibitem{hamaguchi} S. Hamaguchi and S. M. Rossnagel, J. Vac.
Sci. Technol. {\bf B 13}, 183 (1995). 

\bibitem{westwood}
W. D. Westwood, in {\it Microelctronic Materials and Processes},
R. A. Levy ed. (Kluwer, Dordrecht 1989), p.133.

\bibitem{simbad}
S. K. Dew, T. Smy, and M. J. Brett, IEEE Trans. Electron. Dev. {\bf
39}, 1599 (1992).

\bibitem{asym-note}
This is due in part to the intrinsic asymmetry produced by the choice
of the {\it  closest}  highest-coordination site among those within the
$d_{\rm max}$ circle.

\bibitem{methfessel}
It is known (see e.g. M. Methfessel, D. Hennig, and M. Scheffler,
Appl. Phys. A {\bf 55}, 442 (1995), and references therein) 
that the binding energy of metals in a structure
with local coordination $C$ is given 
semiquantitatively  by $E_{\rm bind} = A\, C + B \sqrt{C}$, 
with $A$ and $C$ constants.

\bibitem{dullini} E. Dullini, Nucl. Instr. and Meth. {\bf B 2},
610 (1984). 

\bibitem{bs}
A. Barabasi and H. E. Stanley, {\it Fractal Concepts in Surface Growth},
(Cambridge UP, Cambridge 1995).

\bibitem{friedrich} L. J. Friedrich, S. K. Dew, M. Brett, and T.
Smy, Thin Solid Films  {\bf 226}, 83 (1995);
 S. S. Winterton, T. Smy, S. K. Dew, and
M. J. Brett, J. Appl. Phys. {\bf 78}, 3572 (1995);
 L. J. Friedrich, D. S. Gardner, S. K. Dew, M.
J. Brett, and T. Smy, J. Vac. Sci. Technol. {\bf B 15}, 1780
(1997). 

\end{thebibliography}
\end{document}